\hsize=6in
\vsize=8,5in
\baselineskip=7mm

\tolerance=1000
\looseness=1
\parindent=8mm
\nobreak
\nopagenumbers
\font\cal=cmsy10   

\font\rm=cmr10
\font\rms=cmr7

\font\bfes=cmbx10 scaled\magstep2
\font\it=cmti10
\font\bf=cmbx10 
\font\sl=cmsl10

\def\ln{\mathop{\hbox{\rm ln}}} 

\magnification=1200
{
\parskip 2mm
\noindent {\bfes 
Critical exponents of surface-interacting }

\noindent{\bfes  self-avoiding walks on a  family }

\noindent{\bfes  of truncated n-simplex lattices}

\vskip 0.3cm
\par
}
{
\noindent{\bf Sun\v cica Elezovi\'c--Had\v zi\'c  and Milan
Kne\v zevi\'c}

\noindent {\sl Faculty of Physics, University of
Belgrade, P.O.Box 368, 11001 Belgrade, Serbia, Yugoslavia}

\par
}
\vskip 0.5cm  
{\bf Abstract:} We study the critical behavior of surface-interacting
self-avoiding random walks on a class of truncated simplex lattices, which
can be labeled by an integer $n\ge 3$. Using the exact renormalization
group method we have been able to obtain the exact values of various
critical exponents for all values of $n$ up to $n=6$. We also derived
simple formulas which describe the asymptotic behavior of these exponents
in the limit of large $n$ ($n\to\infty$). In spite of the fact that the
coordination number of the lattice tends to infinity in this limit, we
found that the most of the studied critical exponents approach certain
finite values, which differ from corresponding values for simple random
walks (without self-avoiding walk constraint). 


\vfill
\rightline{Typeset using Plain
{T\kern-.1667em\lower.5ex\hbox{E}\kern-.125em X}} 
\eject

\footline={\hss \rm \folio\hss}
\pageno=2

\parskip=0.4cm
\noindent{\bf 1. Introduction}

\noindent
Configurational properties of a single polymer chain in the vicinity of an
attractive impenetrable wall has been studied long time ago (see, for
instance, [1]), as a problem of great theoretical and practical 
importance. The general picture that springs from these studies [2-4] 
reveals that, under certain conditions, polymer chain can undergo an
adsorption-desorption transition. The essential physics of a polymer chain
near an surface can be captured by the self-avoiding random
walk (SAW) model on a semi-infinite lattice, with an energy contribution
$\epsilon_w$ for each step (monomer) of the walk along the lattice boundary. This
leads to an increased probability, characterized by the Boltzmann factor
$w=\exp(-\epsilon_w/k_BT)$, of making a step along the attractive wall
($\epsilon_w<0$, $w>1$ for any finite temperature $T$). For low
temperatures, the polymer chain is basically pinned at the substrate, while
at higher temperatures all polymer configurations have almost same weights
and a nonadsorbed behavior prevails. The transition between these two
regions is marked by a critical adsorption temperature $T_{\hbox{\rms
a}}$, with a desorbed phase for $T>T_{\hbox{\rms a}}$ and an adsorbed
phase for $T<T_{\hbox{\rms a}}$. The asymptotic behavior of the average
number $M$ of steps of the walk along the boundary can be summarized in
the following way [4] 
$$ M\sim
\left\{\matrix{ N (T_{\hbox{\rms a}}-T)^{1/\Phi -1},&  T<T_{\hbox{\rms a}}
\cr N^\Phi ,& T=T_{\hbox{\rms a}}    \cr (T-T_{\hbox{\rms a}})^{-1} ,&
T>T_{\hbox{\rms a}} \cr}\right.
\, , \eqno(1.1)
$$
where $N$ denotes average number of monomers, and $\Phi$ is the 
crossover exponent.  A more complete description of the statistics of SAWs
near a surface requires a knowledge of the asymptotic behavior of the
numbers ${\hbox{\cal C}}_1(N,T)$, 
${\hbox{\cal C}}_{11}(N,T)$ and 
${\hbox{\cal C}}_s(N,T)$ of different polymer
configurations with one, both and no ends at the wall, respectively. It is
generally believed [4] that these numbers as $N\to\infty$ follow the
asymptotic laws 
$${\hbox{\cal C}}_1(N,T)\sim\mu^NN^{\gamma_1-1}\>,\quad
{\hbox{\cal C}}_{11}(N,T)\sim\mu^NN^{\gamma_{11}-1}\>, \quad
{\hbox{\cal C}}_s(N,T)\sim\mu^NN^{\gamma_s-1}\>,\eqno(1.2)$$
where $\mu=\mu(T)$ is a continuous function of temperature, and $\gamma_1$,
$\gamma_{11}$ and $\gamma_s$ are the associated critical exponents. It
turns out that these exponents assume distinct values in different
temperature regions [5]. 

The most of theoretical efforts so far have been devoted to a study of suitable
models of surface-interacting SAWs on standard homogeneous spaces.
Recently, a considerable research activity have appeared in the study of
SAWs placed on fractal spaces [6-11], as models of polymers in
nonhomogeneous environment. Aside from being interesting in its own right,
we believe that such studies may yield some insights into more difficult
questions related to the behavior of polymers in disordered systems. 
In the present paper we exhibit the large-scale behavior of
surface-interacting SAWs on truncated $n$-simplex lattices [12], which
provide a whole family of fractals that can be labeled by an integer $n\ge
3$. Fractal dimension of these latices 
increases indefinitely when $n\to\infty$, while their spectral dimension
remain limited to a relatively narrow region (see (2.1)). 
Using the renormalization-group (RG) method we 
obtain the exact values of the critical exponents $\Phi$, $\gamma_1$,
$\gamma_{11}$, and $\gamma_s$ in the case of SAWs on 
$4$-, $5$- and $6$-simplex lattice. Then, applying an extension of the
approach described in refs. [13-14], we have been able to determine the
limiting behavior of these exponents, valid for large values of $n$. It is
intersting to note here that the limiting values of
$\gamma_1$, $\gamma_{11}$, and $\gamma_s$ do not coincide with the
corresponding values of these exponents for an analogous random-walk
problem [15], in spite of the fact that coordination number of these
lattices tends to infinity for $n\to\infty$.

The paper is organized as follows. In section 2. we present our approach
and calculate the exact and asymptotic values of the crossover exponent
$\Phi$. This method is then extended to the case of open SAWs, in section
3, where we also display the exact values and asymptotic behavior of  
the exponents $\gamma_1, \gamma_{11}$ and $\gamma_s$. Some final remarks
and an overall discussion we give in section 4.

\eject

\noindent{\bf 2. Crossover exponent $\Phi$}

\noindent
The family of truncated $n-$simplex lattices has been introduced by
Dhar [12],  who also developed simple RG approach to describe
critical behavior of various statistical-mechanical models on these
lattices. The lattices are defined recursively: One starts with a
complete graph of $n$ points and replaces each of these points by a new
complete graph of $n$ points (see Fig. 1 for the case $n=6$). 
The subsequent stages are constructed self-similarly, by repeating this
procedure, so that the complete lattice is obtained in the limit when the
number of such iterations tends to infinity. The fractal $\bar d$ and spectral
$\tilde d$ dimensions of these lattices are given as 
$$\bar d={{\ln n}\over{\ln 2}}\,, \qquad
\tilde d={{2\ln n}\over{\ln(n+2)}} \, , 
\eqno(2.1)$$
respectively. In the case under study, it is assumed that the 
adsorption boundary of a truncated $n-$simplex is a truncated
$(n-1)-$simplex lattice.  The fractal $\bar d_s$
and the spectral $\tilde d_s$ dimension of the adsorption surface can be
presented by (2.1), where $n$ has to be replaced by $(n-1)$.

The crossover exponent $\Phi$ for the SAW model was found for
$n=3, 4$ [6], and $n=5$ [8]. Here we shall present a general framework
of the RG method for studying surface-interacting SAWs on the family of
$n-$simplex fractals on the example of the $6-$simplex lattice.

\noindent{\bf (a) Exact renormalization group approach}

\noindent
In order to study influence of the adsorbing wall on the SAW statistics,
in addition to $w=\exp({-\epsilon_w/k_BT})$, we 
introduce the Boltzmann factor $t=\exp({-\epsilon_t/k_BT})$, where
$\epsilon_t$ denotes the energy of a monomer that appears in the layer 
adjacent to the wall. Here we should set $\epsilon_t>0$
so as to prevent the tendency of a SAW towards being always
adsorbed [6]. We assign the weight $x$ to each step in the bulk (away
from the wall), the weight $wx$ to each step on the adsorbing wall, and
the weight $tx$ to each step in the layer adjacent to the wall. To
calculate the critical exponent $\Phi$, it is necessary to consider  
only a finite number of restricted partition
functions, which are defined as  weighted sums over 
all never-starting and never-ending SAWs for the given stage $r$ 
of the iterative construction of the lattice. 
It is not difficult to see that in the case of a $6-$simplex lattice 
there are eight such restricted partition functions,
 $A^{(r)}$, $A_1^{(r)}$, $A_2^{(r)}$, $B^{(r)}$, $B_1^{(r)}$, $B_2^{(r)}$, $C^{(r)}$ and
$C_1^{(r)}$,  which are depicted in Fig. 2.  
The recursive nature of the fractal under consideration implies
the following general form of the recursion relations:
$$
\eqalignno{A'=\,&\sum_{N_A}\sum_{N_B}\sum_{N_C}
A_{N_A,N_B,N_C}
A^{N_A}B^{N_B}C^{N_C}\, ,&(2.2a)\cr
B'=\,&\sum_{N_A}\sum_{N_B}\sum_{N_C}
B_{N_A,N_B,N_C}A^{N_A}B^{N_B}C^{N_C}\, ,&(2.2b)\cr
C'=\,&\sum_{N_A}\sum_{N_B}\sum_{N_C}
C_{N_A,N_B,N_C} A^{N_A} B^{N_B} C^{N_C}\, ,&(2.2c)\cr
A_1'=\,&{{{  A_1}}^2} + 3\,{{{  A_1}}^3} + 6\,{{{  A_1}}^4} + 6\,{{{  A_1}}^5} + 
 12\,{{{  A_1}}^3}\,{  B_1} + 
  30\,{{{  A_1}}^4}\,{  B_1} +  
 18\,{{{  A_1}}^2}\,{{{  B_1}}^2} + 
  78\,{{{  A_1}}^3}\,{{{  B_1}}^2}\cr
& + 
  96\,{{{  A_1}}^2}\,{{{  B_1}}^3} + 132\,{  A_1}\,{{{  B_1}}^4} + 
  132\,{{{  B_1}}^5} +   F_{A_1A}\,A+F_{A_1B}\,B+F_{A_1C}\,C&(2.3a)\cr
A_2'=\,&   F_{A_2A}\,A+F_{A_2B}\,B+F_{A_2C}\,C&(2.3b)\cr
B_1'=\,&{{{  A_1}}^4} + 2\,{{{  A_1}}^5} +
 4\,{{{  A_1}}^3}\,{  B_1} + 
  13\,{{{  A_1}}^4}\,{  B_1} + 
  32\,{{{  A_1}}^3}\,{{{  B_1}}^2} + 88\,{{{  A_1}}^2}\,{{{  B_1}}^3}
 + 
  22\,{{{  B_1}}^4}\cr
& + 220\,{  A_1}\,{{{  B_1}}^4} + 186\,{{{  B_1}}^5} + 
  F_{B_1A}\,A+F_{B_1B}\,B+F_{B_1C}\,C\, ,&(2.3c)\cr
B_2'=\,&
  F_{B_2A}\,A+F_{B_2B}\,B+F_{B_2C}\,C\, ,&(2.3d)\cr
C_1'=\,&  F_{C_1A}\,A+F_{C_1B}\,B+F_{C_1C}\,C\, ,&(2.3e)\cr}
$$
where we have used the prime symbol as a superscript for
the $(r+1)$-th order functions and no indices for the
$r$-th order functions. The coefficients $A_{N_A,N_B,N_C}$,
$B_{N_A,N_B,N_C}$ and $C_{N_A,N_B,N_C}$ are numbers of ways
in which the corresponding part of the SAW path in the
bulk, within an $(r+1)-$th stage fractal structure, can be comprised of
the SAW paths within the fractal structures of the next lower order.
The coefficients $F$ are polynomials in $A_1, A_2, B_1$ and $B_2$, and
they are given in the Appendix, together with  the
explicit forms of the relations $(2.2)$.

The above set of relations $(2.2)$ and $(2.3)$ can be considered as a set
of RG equations. The bulk relations $(2.2)$ are independent of the
recursions (2.3) for surface partition functions, and they have already been
analysed in [13]. It turns out that system (2.2) has several
nontrivial  fixed points, but only one of them can be reached starting
with the SAWs initial conditions
$$\eqalignno{
A^{(1)}&=\,x+4x^2+12x^3+24x^4+24x^5\, ,\cr
B^{(1)}&=\,x^2+4x^3+6x^4\, ,\cr
C^{(1)}&=\,x^3\, .&(2.4)\cr}
$$
That happens for the critical value of the fugacity,
$$
x=x_c={1\over\mu}\approx 0.137359    \, ,  \eqno(2.5)
$$
and the pertinent fixed point is given by
$$
A^*\approx 0.262352\,, \quad B^*\approx 0.017588\,, 
\quad C^*\approx 0.000701\, .
\eqno(2.6)
$$
In accord with the accepted physical picture about the interaction
parameters, we assume the following initial conditions for the 
remaining RG parameters
$$\eqalignno{A_1^{(1)}&=\,xw(1+3xw+6x^2w^2+6x^3w^3)
+x^2t^2(1+6xw+18x^2w^2+24x^3w^3)\,, \cr
A_2^{(1)}&=\,xt(1+4xw+12x^2w^2+24x^3w^3+24x^4w^4) \, ,\cr
B_1^{(1)}&=\,x^2w^2(1+2xw)+2x^3wt^2(1+3xw) \,  ,\cr
B_2^{(1)}&=\,x^2wt(1+4xw+6x^2w^2)  \,  ,\cr
C_1^{(1)}&=\,x^3w^2t \, . &(2.7)\cr}
$$
The numerical study of the relations $(2.2)$ and $(2.3)$ with the
initial conditions $(2.4)$ and $(2.7)$ shows that for any fixed 
value of $t<1$ there are three different temperature regions. We are
going to discuss them separately, starting with the high
temperature region.

(i) At high temperatures, that is, for
$w<w^*(t)$ (here $w^*(t)$ denotes a critical $t$~--~dependent value of
$w$), the critical fugacity is constant and equal to its bulk critical
value (2.5). For all these values of temperature  the
bulk SAW fixed point is reached
$$
(A^*,\,A_1^*,\,A_2^*,B^*,B_1^*,B_2^*,C^*,C_1^*)
=(A^*,\,0,\,0,\,B^*,\,0,\,0,\,C^*,\,0\,)\, . \eqno(2.8)
$$
The fraction of SAW steps in contact with the surface vanishes in
this temperature region, so that the polymer is in the desorbed state.
Linearization about this fixed point leads to only one relevant 
eigenvalue
 $$\lambda_\nu \approx\,3.52148 \, ,\eqno(2.9)
$$
which yields the value of the end--to--end distance
critical exponent $\nu=\ln 2/\ln\lambda_\nu\approx 0.55061$.

(ii) When the temperature is lowered an adsorption transition
occurs for $w=w^*(t)$. In that case $x_c(w^*)$ is still equal 
to its bulk value $(2.5)$, but equations $(2.2)$ and $(2.3)$ iterate 
towards the symmetric fixed point
$$
(A^*,\,A_1^*,\,A_2^*,B^*,B_1^*,B_2^*,C^*,C_1^*)
=(A^*,\,A^*,\,A^*,\,B^*,\,B^*,\,B^*,\,C^*,\,C^*\,)\, ,  \eqno(2.10)
$$
where a balance between the attractive polymer--surface potential and an 
effective "entropic" repulsion sets in ("special transition" in the
common language of surface phase transition).
This fixed point has two relevant eigenvalues: 
$\lambda_\nu$, given by (2.9), and 
$$
\lambda_\Phi\approx \, 2.97273   \, . \eqno(2.11)
$$
The larger of them determines the end--to--end distance critical exponent
$\nu$, which means that this index has the same value  as in 
the high temperature region. On the other hand, it is known [16] that the
crossover exponent $\Phi$ involves both eigenvalues: 
$$
\Phi=\, {{\ln\lambda_\Phi}\over{\ln\lambda_\nu}}\approx 0.86544 \, .
\eqno(2.12)
$$

(iii) In the low temperature region, $w>w^*(t)$, the critical
fugacity $x_c(w)$ is a decreasing function of $w$, while
recursion relations $(2.2)$ and $(2.3)$ iterate towards
the fixed point
$$
(A^*,\,A_1^*,\,A_2^*,B^*,B_1^*,B_2^*,C^*,C_1^*)
=(0,\,(A^*)_{n=5},\,0,\,0,\,(B^*)_{n=5},\,0,\,0,\,0\,)\, ,  
\eqno(2.13)
$$
where $((A^*)_{n=5}\approx 0.326491, (B^*)_{n=5}\approx 0.027930)$ 
is the polymer bulk fixed point for the case of a $5-$simplex fractal [13]. 
The above fixed point describes the critical properties of a polymer chain in 
adsorbed state. It is clear, therefore, that an adsorbed polymer  
on the $6-$simplex fractal has the same properties as a polymer chain in the
bulk of a $5-$simplex lattice.

 It is not feasible to pursue
the above described approach for large values of $n$, because the 
number of required partition functions, and especially their
complexity, grow up rather quickly with $n$. To overcome this
difficulty, we shall apply here an approximate procedure which provides a
direct extension of a similar technique used earlier in the study of
the bulk critical behavior of SAWs on the same class of lattices
[13, 14]. Although one could expect that such an approach should be valid only
in the asymptotic region $n\to\infty$, it turns out that this method gives
quite accurate values of $\Phi$ even for moderate values of $n$ ($n\sim 5$).

\medskip

\noindent
{\bf (b) An approximate approach for the critical exponent $\Phi$}

\noindent
An analysis of corresponding  recursion relations reveals 
that critical behavior of a polymer chain, in the limit $n\gg 1$, can be
described in terms of a small number of restricted partition functions. For
example, it has been established [13] that the position of the bulk fixed
point takes the asymptotic form (compare (2.6) and see Fig. 2)
$$
A^*\sim {1\over n}\,, \quad
B^*\sim {1\over n^2}\,, \quad
C^*\sim {1\over n^3}\,, \> \cdots \, .\eqno(2.14)
$$
In the neighborhood of this fixed point, one can simplify the
exact recursion relations by neglecting 
the variables $B$, $C$, $\cdots$, which approach zero much faster than $A$
when $n\to\infty$. In this way, the bulk critical behavior of a polymer
chain can be described by using only one variable, which satisfies the
simple recursion relation [13]
$$
A'=\, \sum_{k=0}^{n-2}k! \left({{n-2}\atop k}\right) A^{k+2} \, ,
\eqno(2.15)
$$

In a similar way, in order to calculate the critical exponent $\Phi$ 
($n\gg 1$) one can keep only the partition functions $A$,
$A_1$, and $A_2$ and neglect all the others (see Fig. 2).
Then, in addition to the above formula, one has to consider the recursion
relations for surface variables 
$$\eqalignno{
A_1'=\, &\sum_{k=1}^{n-2}(k-1)!\left({{n-3}\atop{k-1}}\right) A_1^{k-1}
\left(A_1^2+k A A_2^2\right) \, , &(2.16a)\cr
A_2'=\, &\sum_{k=0}^{n-2}k! \left({{n-2}\atop k}\right) A A_2 A_1^k \, .
&(2.16b)\cr}
$$
Indeed, critical exponent $\Phi$ is still determined by relation (2.12)
with $\lambda_\nu$ and $\lambda_\Phi$ being two largest eigenvalues 
associated to the symmetric fixed point $A=A_1=A_2=A^*$, 
$$
\lambda_\nu=\,n+1-{1\over{A^*}} \,, \eqno(2.17)
$$
and
$$
\lambda_\Phi = {5 - n- A^*(1  + 3n - n^2) + \sqrt{\Delta}
       \over{2A^*(n-2)}}\,,\eqno(2.18a)
$$
with
$$
\Delta={9 - 2n+n^2 - 2 A^*(1+4n-2n^2+n^3) - {A^*}^2(7 - 10n
+n^2+2n^3-n^4)}\,,\eqno(2.18b) 
$$
where $A^*$ denotes the fixed point of (2.15). For some specific values of
$n$ we have determined this fixed point and the corresponding eigenvalues
numerically. In this way we
calculated the values of the crossover critical exponent which are
presented in Fig. 3. For the sake of comparison we also presented (Fig.
3a) the exact values of this exponent for the case $n=3,4,5$ [6], [8]
and $n=6$ (see (2.12)). One can see that the exact and corresponding
approximate value of $\Phi$ are very close to each other even for these
moderate values of $n$. One can also notice that for larger values of $n$ 
exponent $\Phi$ approaches, almost linearly in $1/n \ln n$, values that
are very close to 1, (see Fig 3b). This observation can be further
corroborated, by using the earlier established [14] asymptotic formula for the
position of the bulk fixed point
$$
A^*\sim (1+s)/(n+1),  \quad  s= \sqrt{(\ln n)/n}\,,\quad n\gg 1.\eqno(2.19)
$$
Indeed, the above formulas lead to the following estimate
$$
\Phi\sim 1- {1\over{n\ln n}}, \quad n\gg 1,\eqno(2.20)
$$
which is in excellent agreement with our numerical results presented in
Fig. 3b. The fact that critical exponent $\Phi$ tends to 1 as $n\to\infty$
is not surprising, since in this limit the number of sites on the adsorbing
surface becomes comparable to the total number of the available sites on
the lattice.

It is also of some interest to compare our results with the upper
($\Phi_u$) and lower ($\Phi_l$)
bounds on the crossover exponent, proposed recently [6] by Bouchaud and
Vannimenus for SAWs on fractals. One can observe (see Fig. 3) that both
exact and approximate values of $\Phi$ satisfy their bounds
$$
\Phi_l=1-(\bar d-\bar d_s)\nu \le \Phi\le
\Phi_u=\bar d_s/\bar d,  \eqno(2.21)
$$
with $\bar d$ and $\bar {d_s}$ being the fractal dimension of the lattice
and the adsorbing wall, respectively. Besides, it is interesting to note
here that for large values of $n$ the approximate values of $\Phi$ are almost
equal to the values of $\Phi_l$ (see Fig. 3b). On the other hand, it was
shown recently [15] that the lower bound $\Phi_l$ presents, in fact, the
exact value of the crossover exponent for the simple random walks (without
SAWs constraint) on the same class of lattices. This is in accordance with
the expectation that in the limit of large coordination number (i.e.
large $n$) statistics of SAWs should be similar to the one for random walks. 

\bigskip

\noindent{\bf 3. Critical exponents $\gamma_1$, $\gamma_{11}$ and
$\gamma_s$}

\noindent
To calculate the critical exponents $\gamma_1$, $\gamma_{11}$ and
$\gamma_s$ it is helpful to introduce the convenient generating
functions
$$
{\hbox{\cal C}}_1(x,T)=
\sum_{N=1}^\infty {x^N}\sum_{M,L=1}^N {{\hbox{\cal C}}_1(N,M,L) w^Mt^L}=
\sum_{N=1}^\infty {{\hbox{\cal C}}_1(N,T) x^N}\>,\eqno(3.1a)
$$
$${\hbox{\cal C}}_{11}(x,T)=
\sum_{N=1}^\infty{x^N}\sum_{M,L=1}^N {{\hbox{\cal C}}_{11}(N,M,L) 
w^Mt^L}=
\sum_{N=1}^\infty {{\hbox{\cal C}}_{11}(N,T) x^N} 
\>,\eqno(3.1b)
$$
$$
{\hbox{\cal C}}_s(x,T)=\sum_{N=1}^\infty {x^N}\sum_{M,L=1}^N 
{{\hbox{\cal C}}_s(N,M,L)
w^Mt^L}=\sum_{N=1}^\infty {{\hbox{\cal C}}_s(N,T) x^N} \>,\eqno(3.1c)
$$
where ${\hbox{\cal C}}_1(N,M,L)\ 
\bigl({\hbox{\cal C}}_{11}(N,M,L)\bigr)$ represents the 
averaged number of $N$-step SAWs with $M$ steps on the surface,
and $L$ steps in the layer adjacent to the wall provided one
(both) end(s) of the walk is (are) anchored to the wall,
while ${\hbox{\cal C}}_s(N,M,L)$ is the averaged number of SAWs with no ends anchored to
the wall. The weighting factors $x, w$ and $t$ were defined in the
foregoing section. Assuming that numbers ${\hbox{\cal C}}_1(N,T)$, 
${\hbox{\cal C}}_{11}(N,T)$
and ${\hbox{\cal C}}_s(N,T)$ behave as in (1.2),
the leading singular behavior of the generating functions, 
when $x$ approaches $x_c=1/\mu(T)$ from below, is of the form
$$
{\hbox{\cal C}}_1(x,T)\sim (1-x\mu )^{-\gamma_1}\, ,\quad
{\hbox{\cal C}}_{11}(x,T)\sim (1-x\mu )^{-\gamma_{11}}\, ,\quad
{\hbox{\cal C}}_s(x,T)\sim (1-x\mu)^{-\gamma_s}\, . \eqno(3.2)
$$
The above generating functions can be expressed in terms of a finite
number of restricted partition functions [7]. In addition to 
the restricted partition functions defined in the preceding section,
one should introduce restricted partition functions describing
SAW's ending (with one or both ends) somewhere inside of an  
$r-$th stage fractal lattice. As in the case of the 
critical exponent $\Phi$, we shall first outline the general method
for calculating the  exponents $\gamma_1$, $\gamma_{11}$,
and $\gamma_s$  on the $6-$simplex lattice example.

\eject

\noindent{\bf (a) Exact method for calculating 
$\gamma_1$, $\gamma_{11}$ and $\gamma_s$}

To treat the statistics of open SAWs, one has to introduce a 
number of additional restricted partition functions. In the case $n=6$,
for example, a complete description of the above generating
functions (3.1) needs 46 restricted partition
functions, in addition to those presented in Fig. 2. Let us start with the
simplest one --$\,{\hbox{\cal C}}_{11}$, 
which can be expressed in terms of 15 functions $D_1$, $D_2$,
$E_1$, $E_2$, $E_3$, $F_1$, $F_2$, $F_3$, $G_1$, $G_2$, $H_1$, $H_2$,
$H_3$, $I_1$, and $I_2$, depicted in Fig. 4.
These 15 functions describe different SAW configurations
with ends at the vertices on the adsorbing wall.
It is not difficult to show that ${\hbox{\cal C}}_{11}$ can be 
written in the form
$$
{\hbox{\cal C}}_{11}(x,T)=\sum_{r=1}^{\infty}{{F_{11}^{(r)}}\over{5^r}}\>,
\eqno(3.3)
$$
where $F_{11}^{(r+1)}$ is a  quadratic function in $D_1^{(r)}$, $D_2^{(r)}$,
$E_1^{(r)}$, $E_2^{(r)}$, $E_3^{(r)}$, 
$F_1^{(r)}$, $F_2^{(r)}$, and $F_3^{(r)}$, and linear in 
$G_1^{(r)}$, $G_2^{(r)}$, $H_1^{(r)}$, $H_2^{(r)}$, 
$H_3^{(r)}$, $I_1^{(r)}$, and $I_2^{(r)}$,  
with coefficients being polynomials in $A^{(r)}$, $A_1^{(r)}$, 
$A_2^{(r)}$, $B^{(r)}$, $B_1^{(r)}$, $B_2^{(r)}$, $C^{(r)}$, and
$C_1^{(r)}$ (see, e.g., [7] for an explicit construction). 
For arbitrary $r$, the self-similarity
of the $6-$simplex lattice imply a recursion relation of the type
$$
\left(
\matrix{D_1^{(r+1)}\cr
D_2^{(r+1)}\cr
E_1^{(r+1)}\cr
E_2^{(r+1)}\cr
E_3^{(r+1)}\cr
F_1^{(r+1)}\cr
F_2^{(r+1)}\cr
F_3^{(r+1)}\cr}
\right)={\hbox{\bf \^ A}}
\left(
\matrix{D_1^{(r)}\cr
D_2^{(r)}\cr
E_1^{(r)}\cr
E_2^{(r)}\cr
E_3^{(r)}\cr
F_1^{(r)}\cr
F_2^{(r)}\cr
F_3^{(r)}\cr}
\right)\, , \eqno(3.4)
$$ 
 where elements of the matrix ${\hbox{\bf \^ A}}$ are some polynomials
in $A^{(r)}$, $A_1^{(r)}$, $A_2^{(r)}$, $B^{(r)}$, $B_1^{(r)}$,
$B_2^{(r)}$, $C^{(r)}$, and $C_1^{(r)}$.  The recursion relations (2.2)
and (2.3), together with the matrix relation (3.4), and an analogous
relation involving variables $G_1$, $G_2$, $H_1$, $H_2$, $H_3$, $I_1$ and
$I_2$, form a closed set of the recursion relations.  Starting with the
appropriate initial conditions for the restricted partition functions, it
is possible to determine the value of the generating function
${\hbox{\cal C}}_{11}(x,T)$ for arbitrary values of $x$ and $T$. 
As we are interested here only in the asymptotic form of 
${\hbox{\cal C}}_{11}(x,T)$, the precise forms of
the recursion relations for $G_1$, $G_2$, $H_1$, $H_2$, $H_3$, $I_1$ and
$I_2$, as well as the explicit form of corresponding initial conditions,
are not necessary, so we do not give them here. The elements of the matrix
${\hbox{\bf \^ A}}$ are very cumbersome polynomials, so we do not present
them either, but they are available upon request.

When $x$ is sufficiently close to $x_c$, so that $x_c-x\ll\varepsilon\ll
1$ ($\varepsilon>0$), it turns out that under iterations, for
$r<r_0=\ln(\varepsilon/(x_c-x))/\ln\lambda_\nu\gg 1$, each restricted
partition function either tends to certain finite constant
or follows the simple power law $\sim\lambda^r$ (with a specific value of 
$\lambda$ for each particular function, in general).
In particular, this holds for the correlation functions $A^{(r)}$,
$A_1^{(r)}$, $A_2^{(r)}$, $B^{(r)}$, $B_1^{(r)}$, $B_2^{(r)}$, $C^{(r)}$ and
$C_1^{(r)}$, which present some nonincreasing functions of the iteration
index $r$. Taking into account that $F_{11}^{(r)}$ is a quadratic function
in all the other partition functions, we conclude that the summand in
(3.3) follows the power law $(\lambda_{11}^2/5)^r$, where $\lambda_{11}$
describes the way in which diverges the dominant of 
the functions
$
D_1^{(r)}, D_2^{(r)},
 E_1^{(r)},  E_2^{(r)},  E_3^{(r)}, 
 F_1^{(r)},  F_2^{(r)},  F_3^{(r)},  G_1^{(r)},  G_2^{(r)},
 H_1^{(r)},  H_2^{(r)},  H_3^{(r)},  I_1^{(r)},$ and  $I_2^{(r)}$.
For $r>r_0$ all these partition functions rapidly
approach some constants, so that corresponding summands in (3.3) becomes
negligible. Thus, the major contribution to the sum in (3.3) 
comes from the term with $r\approx r_0$, that is
$$
{\hbox{\cal C}}_{11}(x,T)\sim (\lambda_{11}^2/5)^{r_0} \, . \eqno(3.5)
$$
This enables us to express the critical exponent $\gamma_{11}$ in 
terms of $\lambda_{11}$ and $\lambda_\nu$,
$$
\gamma_{11}={{\ln{{\lambda_{11}^2}\over 5}}\over{\ln\lambda_\nu}}
\, . \eqno(3.6)
$$

An examination of the recursion relations (3.4) in the vicinity
of relevant fixed points 
\line{leads to the following:\hfill} 
\parskip 0mm
\vskip 0mm
(i) In the high temperature region, controlled by the bulk fixed
point (2.8), the largest contribution to the function $F_{11}^{(r)}$ comes
from the function $D_2$, which, for $r<r_0$, behaves as
$$
D_2^{(r)}\sim \,(5A^*)^r=\lambda_{11}^r\approx (1.31176)^r
  \, , \eqno(3.7)
$$
so that we find
$$
\gamma_{11}\approx\, -0.8473              \, . \eqno(3.8)
$$
\vskip 0mm
(ii) In the case of the adsorption transition, the relevant fixed point is
the symmetric one (see (2.10)), and $\lambda_{11}\approx \,5.21757 $
is the largest eigenvalue of the matrix ${\hbox{\bf \^ A}}$,
which appears in the recursion relation (3.4), leading to 
$$
\gamma_{11}\approx \,1.3461            \, . \eqno(3.9)
$$
\vskip 0mm
(iii) In the low temperature region SAW is mostly adsorbed, so 
the critical exponent $\gamma_{11}$ is equal to the bulk $\gamma$ exponent 
for SAW on a $5-$simplex lattice [13], i.e. $\gamma_{11}\approx\,
1.4875$. 
\parskip 2mm

The evaluation of the critical exponent $\gamma_1$ goes 
along the same lines, except for the fact that one should introduce
some additional restricted partition functions. These new functions 
are presented in Fig. 5, and they describe the SAW configurations that have
one end somewhere in the bulk. One can show that the final formula 
for $\gamma_1$ takes the form
$$
\gamma_1=\,{{\ln(\lambda_{11}\lambda/5)}\over{\ln\lambda_\nu}} \, ,
\eqno(3.10)
$$
where $\lambda$ denotes the largest eigenvalue of the corresponding matrix
appearing in the recursion relations involving bulk variables $D$, $E$ and
$F$.  
Numerically we find: $\,\gamma_1\approx 0.39499$ for the desorbed SAW,
  $\gamma_1\approx  1.49173$  for the SAW at the adsorption transition, 
and $\gamma_1\approx 1.4875$ for an adsorbed polymer chain (which is equal
to the value of $\gamma$ for the SAW on the $5-$simplex lattice).     
In a similar way, we find
$$
\gamma_s=\, {{\ln(\lambda^2/5)}\over{\ln\lambda_\nu}} \, ,
\eqno(3.11)
$$
which means that $\gamma_s$ has the same value at the point of the
adsorption transition as well as in the high temperature phase 
($\gamma_s\approx 1.63732$ for $T\ge T_{\hbox{\rms a}}$, while in the low 
temperature region its value coincides with the value of 
$\gamma(n=5)\approx 1.4875$). 

The above described approach can be, in principle, applied for general
$n-$ simplex lattice. In particular, it is easy to see that critical
exponents $\gamma_1$, $\gamma_{11}$ and $\gamma_s$ can be expressed in
terms of the above mentioned eigenvalues $\lambda_{11}$, $\lambda_\nu$ 
and $\lambda$,
$$
\gamma_{11}={{\ln{{\lambda_{11}^2}\over{n-1}}}\over{\ln\lambda_\nu}}\, ,
\quad 
\gamma_1=\,{{\ln{{\lambda_{11}\lambda}\over{n-1}}\over{\ln\lambda_\nu}}}
\, , \quad
\gamma_s={{\ln{{\lambda^2}\over{n-1}}\over{\ln\lambda_\nu}}} \, , 
\eqno(3.12)
$$
We have determined all these indices for $4-$simplex and $5-$simplex
lattice (the case $n=3$ has been already studied in some details [7]).
Our results can be summarized in the following way:
for the  SAWs on the $4-$simplex lattice we have obtained
$
\gamma_{11}\approx\,-0.57558 \, ,\>\gamma_1\approx\,0.57514  \, 
,\>\gamma_s\approx\,1.7259 
$
for $T>T_{\hbox{\rms a}}$, and
$
\gamma_{11}\approx\,1.1595 \, ,\>\gamma_1\approx\, 1.4427 \, 
,\>\gamma_s\approx\,1.7259  \,  
$
at $T=T_{\hbox{\rms a}}$, whereas in the case $n=5$ we have found
$
\gamma_{11}\approx \,-0.73826 \, 
,\>\gamma_1\approx\,0.46562  \, 
,\>\gamma_s\approx\,1.6695  \,  
$
for the high temperature phase, and
$
\gamma_{11}\approx\,1.30962 \, ,\>
\gamma_1\approx\,1.4896  \, ,\>
\gamma_s\approx\,1.6695
$, at the adsorption transition.
It is interesting to note here that these values of the critical exponents
satisfy the usual scaling relations for surface interacting SAWs on fractals [7]
$$
\gamma_s=2\gamma_1-\gamma_{11}=\gamma +\nu (\bar d-\bar d_s)
\, , \eqno(3.13)
$$
where the bulk critical exponents $\gamma$ and $\nu$ can be written
in terms of $\lambda$ and $\lambda_\nu$: $\,\gamma
=\ln(\lambda^2/n)/\ln\lambda_\nu$ and $\nu =\ln 2/\ln\lambda_\nu$.

As it has been emphasized, the exact calculation of the critical exponents
$\gamma_1$, $\gamma_{11}$ and $\gamma_s$ becomes very complicated for
larger values of $n$, so that we are going now to present an
asymptotic analysis of these exponents, which is valid for $n\gg 1$.

\noindent{\bf (b) Approximate calculation of the critical exponents
$\gamma_1$, $\gamma_{11}$ and $\gamma_s$}

To treat open walks, in the limit of large $n$, we use here an extension
of the approach proposed in ref. [13]. It turns out that in this limit we
can simplify the problem considerably, and study 
only 5 restricted partition functions: 
$\,D^{(r)}$, $D_1^{(r)}$, $D_2^{(r)}$, $D_3^{(r)}$ and $D_4^{(r)}$, which
describe the SAW configurations with one end  lying somewhere inside
an $r-$th order $n-$simplex. These functions have the same meaning as
those presented in Figs. 4. and 5. for the case of the $6-$simplex
lattices and they obey the following recursion relations
$$\eqalignno{
\left(
\matrix{D_1'\cr   
D_2'\cr  }   \right)=\,&\left(\matrix{
f_{D_1D_1}&f_{D_1D_2}\cr
f_{D_2D_1}&f_{D_2D_2}\cr}
\right)
\left(
\matrix{D_1\cr   
D_2\cr  }   \right)\, , &(3.14a)\cr
\left(
\matrix{D_3'\cr   
D_4'\cr  }   \right)=\,&\left(\matrix{
f_{D_1D_1}&f_{D_1D_2}\cr
f_{D_2D_1}&f_{D_2D_2}\cr}
\right)
\left(
\matrix{D_3\cr   
D_4\cr  }   \right)+D\,\left(
\matrix{f_{D_3D}\cr   
1\cr  }   \right)\, ,&(3.14b)\cr}
$$
where
$$\eqalignno{
f_{D_1D_1}=\,&1+(n-2)A_1+(n-2)(n-3)A_1^2\,+\cdots +(n-2)!A_1^{n-2}
+A\, A_2^2(n-2)(n-3) \cr
&\times\left(
1+2(n-4)A_1+3(n-4)(n-5)A_1^2+\cdots +(n-3)(n-4)!A_1^{n-4}\right)\, , \cr
f_{D_1D_2}=\,&A\,A_2\,(n-2)\left(1+(n-3)A_1+(n-3)(n-4)A_1^2
+\cdots+(n-3)!A_1^{n-3}\right)\,, \cr
f_{D_2D_1}=\,&A\,A_2 \,\left((n-1)(n-2)+(n-1)(n-2)(n-3)A_1+\cdots +
(n-1)!A_1^{n-3}\right)\cr
f_{D_2D_2}=\,&(n-1)A\, ,\cr
f_{D_3D}=\,&A_2\left(1+(n-2)A_1+(n-2)(n-3)A_1^2+
\cdots+(n-2)!A_1^{n-2}\right)\, . &(3.15)\cr}
$$
In addition, we need the recursion relation for the bulk variable $D$,
which follows from (3.14) for $D_1=D_2=D_3=D_4=D$ and $A_1=A_2=A$ (see also
[13]),  
$$
D'=\, D\,\left(1+(n-1)A+(n-1)(n-2)A^2+\cdots +(n-1)!A^{n-1}\right) \, .
\eqno(3.16)
$$

The high temperature fixed point of the  recursion relations (3.14)
has two relevant eigenvalues 
$$
\lambda_{11}=(n-1)A^*,\quad \lambda =n,\eqno(3.17)
$$
where we use the notation of the preceding subsection. Similarly, at the
point of the critical adsorption transition ($T=T_{\hbox{\rms a}}$) one has 
$$
\lambda_{11}=n-1, \quad \lambda =n .\eqno(3.18)
$$
It is a simple matter to obtain the numerical
values of corresponding critical exponents for each $n$, by taking into
account (3.12). The values of all these exponents, obtained within this
approximate approach, together with corresponding exact values (reported in
subsection 3a), are displayed in Fig. 6. As in the case of the critical
exponent $\Phi$, one can notice that the exact and concomitant approximate
values are quite close to each other, even for small values of $n$, which
provides a justification of the above approach. For large $n$
all these exponents tend to certain finite limiting values that can be
determined (see Fig. 7, where we plotted the approximate values of
$\gamma_1$, $\gamma_{11}$ and $\gamma_s$ against $\ln\ln{n}/\ln{n}$). 
Indeed, for $T>T_{\hbox{\rms a}}$, one can derive the simple asymptotic
forms 
$$
\gamma_{11}\sim \,-2\left(1-{{\ln\ln n}\over{\ln n}}\right) \,,\quad
\gamma_1\sim\, {2\over{\sqrt{n\ln n}}}\, ,\quad
\gamma_s\sim \,2\left(1-{{\ln\ln n}\over{\ln n}}\right)\, ,
\eqno(3.19)
$$
which are in good agreement with our numerical findings (see Fig. 7),
obtained via the approximate approach. A
similar conclusion holds for $T=T_{\hbox{\rms a}}$, in which case we have found
$$
\gamma_{11}\sim\,\gamma_1\sim\,
\gamma_s\sim \,2\left(1-{{\ln\ln n}\over{\ln n}}\right).
\eqno(3.20)
$$

\bigskip

\noindent
{\bf 4. Conclusion}

\noindent
In this paper we have studied critical behavior of surface-interacting
SAWs on the family of truncated $n-$simplex lattices. We have calculated
the exact values of the crossover critical exponent $\Phi$ and the surface
susceptibility critical exponents $\gamma_1$, $\gamma_{11}$ 
and $\gamma_s$, in different temperature regions, for low values of the
lattice parameter $n$ (up to $n=6$). These values are then compared with the
corresponding values that we have obtained by using an approximate
approach, which is expected to be valid in the limit of large $n$. It
is shown that our approximate findings are quite close to the 
exact ones, even for these small values of $n$ (see Fig. 3. and Fig. 6). 

We have found that both exact and approximate values of $\Phi$ increase
as $n$ grows, becoming closer to the lower bound $\Phi_l$ (2.21) for
this exponent (Fig. 3a). For larger  $n$ the values of $\Phi$ (calculated
approximately) monotonically approach 1 (Fig. 3b), in the same way as in
the case of an analogous, recently studied [15], problem of surface
interacting random walks. This is in agreement with a na{\"\i}ve
expectation that the critical behavior of SAWs on a lattice should be more
and more similar to the one for simple random walks, when the coordination
number of the underlying lattice grows up indefinitely.
  The limiting
behavior of the surface susceptibility critical exponents for SAWs is,
however, completely different from those for simple random walks (for
$n\to\infty$ and $T=T_{\hbox{\rms a}}$, for example, one has $\gamma_{11}=\gamma_1=\gamma_s=2$ and 
$\gamma_{11}^{RW}=\gamma_1^{RW}=\gamma_s^{RW}=1$ [15], for SAWs and random
walks, respectively). Nevertheless, the usual scaling relations (3.13) are
satisfied in all these cases. 
One can also notice that  inequality 
$\gamma_{11}(T>T_{\hbox{\rms a}})<\gamma_1(T>T_{\hbox{\rms a}})<
\gamma_{11}(T=T_{\hbox{\rms a}})<\gamma_1(T=T_{\hbox{\rms a}})
<\gamma_s(T\ge T_{\hbox{\rms a}})$
 holds for  small, as well as
for larger values of $n$ (see Fig. 6 and Fig. 7). This is quite plausible,
since the SAWs with at least one end attached to the adsorbing wall
certainly outnumbers the SAWs with both ends at the wall, and also
the number of the monomers at the wall is larger at the adsorption
transition ($T=T_{\hbox{\rms a}}$) than in the high temperature region 
($T>T_{\hbox{\rms a}}$), when the SAWs are desorbed (see (1.2)). 
The values of the critical exponent $\gamma_s$ are the same in the
bulk and the crossover region, and these values are larger than the
corresponding values of $\gamma_1$ and  $\gamma_{11}$ in both regions,
which means that, asymptotically, average number of SAWs with no ends
at the adsorbing wall is larger than the number of  SAWs with ends
at the wall. At the adsorption transition
$\gamma_1$ and $\gamma_{11}$ become almost equal to $\gamma_s$,
when $n\to\infty$ (Fig. 7b), monotonically approaching the value 2, predicted 
for the exponent $\gamma$ [14], which is related to
all possible SAWs. On the contrary, for the desorbed SAWs 
($T>T_{\hbox{\rms a}}$) the difference in the values of
$\gamma_1, \gamma_{11}$, and $\gamma_s$ becomes larger when $n$
increases (Fig. 7a), since $\gamma_{11}\to -2$ and $\gamma_1\to 0$,
when $n\to\infty$.

At the end we can conclude that simple approximation of the exact
RG approach for calculating critical exponents of  SAWs 
on $n$--simplex lattices [13] turned out to be very fruitful
in the case of the surface-interacting SAWs. 
Unfortunately, this approximate method is valid only in the case
of SAW model for a diluted polymer solution, 
whereas the exact RG equations 
allow for investigating other interesting features of the polymer chains
situated on fractals, for instance, adsorption and collapse 
transition [6, 8]. The corresponding analysis of the exact RG equations for the
6--simplex lattice, when interaction between the contiguous monomers  
in the bulk is taken into account, will be presented elsewhere in 
the near future.


\vfill
\eject

\noindent{\bf Appendix}

Here we give the explicit forms of the recursion relations
$(2.2a-c)$, as well as the polynomial functions $F$ appearing in the
relations $(2.3a-e)$, for the case of a truncated 6--simplex lattice.
$$\eqalignno{
A'=\,&25008B^4C^2+528B^5+20544B^5C+6576B^6+11328AB^3C^2+528AB^4\cr
&+15264AB^4C+8688AB^5+A^2+36A^2B^2+384A^2B^3+4992A^2B^3C+5544A^2B^4\cr
&+4A^3+24A^3B+312A^3B^2+1728A^3B^2C+2592A^3B^3+12A^4+120A^4C^2\cr
&+120A^4B+480A^4BC+960A^4B^2+24A^5+48A^5C+216A^5B+24A^6\, , &(A1)\cr
B'=\,&94336B^2C^4+76800B^3C^3+22B^4+48160B^4C^2+372B^5\cr
&+23520B^5C+5440B^6+16672AB^3C^2+440AB^4+17120AB^4C\cr
&+6576AB^5+2832A^2B^2C^2+176A^2B^3+5088A^2B^3C+3620A^2B^4\cr
&+4A^3B+64B^3C^2+832A^3B^2C+1232A^3B^3+A^4+26A^4B\cr
&+144A^4BC+324A^4B^2+4A^5+16A^5C+64A^5B+6A^6\, ,&(A2)\cr
C'=\,&541568C^6+94336B^3C^3+43200B^4C^2+14448B^5C\cr
&+2940B^6+6252AB^4C+2568AB^5+1416A^2B^3C\cr
&+954A^2B^4+208A^3B^3+54A^4B^2+6A^5C+12A^5B+A^6\, , &(A3)\cr}
$$

$$\eqalignno{F_{A_1A} =\,&{{{  A_2} }\,^2}\, + 6\,{  A_1} \,{{{  A_2}}^2} + 
  18\,{{{  A_1}}^2}\,{{{  A_2}}^2} + 24\,{{{  A_1}}^3}\,{{{  A_2}}^2} + 
  12\,{  A_1}\,{{{  A_2}}^2}\,{  B_1} + 
  54\,{{{  A_1}}^2}\,{{{  A_2}}^2}\,{  B_1}\cr
& + 
  36\,{  A_1}\,{{{  A_2}}^2}\,{{{  B_1}}^2} + 
  12\,{  A_1}\,{  A_2}\,{  B_2} + 72\,{{{  A_1}}^2}\,{  A_2}\,{  B_2} + 
  144\,{{{  A_1}}^3}\,{  A_2}\,{  B_2} + 
  72\,{  A_1}\,{  A_2}\,{  B_1}\,{  B_2}\cr
& + 
  432\,{{{  A_1}}^2}\,{  A_2}\,{  B_1}\,{  B_2} + 
  456\,{  A_1}\,{  A_2}\,{{{  B_1}}^2}\,{  B_2} + 
  264\,{  A_2}\,{{{  B_1}}^3}\,{  B_2} + 18\,{  A_1}\,{{{  B_2}}^2}\cr
& + 
  126\,{{{  A_1}}^2}\,{{{  B_2}}^2} + 312\,{{{  A_1}}^3}\,{{{  B_2}}^2} + 
  192\,{  A_1}\,{  B_1}\,{{{  B_2}}^2} + 
  1260\,{{{  A_1}}^2}\,{  B_1}\,{{{  B_2}}^2} + 
  132\,{{{  B_1}}^2}\,{{{  B_2}}^2}\cr
& + 
  2244\,{  A_1}\,{{{  B_1}}^2}\,{{{  B_2}}^2} + 
  1908\,{{{  B_1}}^3}\,{{{  B_2}}^2} + 
  48\,{{{  A_1}}^3}\,{  A_2}\,{  C_1} + 
  144\,{{{  A_1}}^2}\,{  A_2}\,{  B_1}\,{  C_1}\cr
& + 
  288\,{{{  A_1}}^3}\,{  B_2}\,{  C_1} + 
  1296\,{{{  A_1}}^2}\,{  B_1}\,{  B_2}\,{  C_1} + 
  2496\,{  A_1}\,{{{  B_1}}^2}\,{  B_2}\,{  C_1}\cr
& + 
  3552\,{{{  B_1}}^3}\,{  B_2}\,{  C_1} + 
  120\,{{{  A_1}}^3}\,{{{  C_1}}^2} + 2832\,{{{  B_1}}^3}\,{{{  C_1}}^2}
\,, &(A4)\cr
F_{A_1B}=\,&6\,{{{  A_2}}^4} + 18\,{  A_1}\,{{{  A_2}}^4} + 
  12\,{{{  A_2}}^4}\,{  B_1} + 36\,{{{  A_2}}^3}\,{  B_2} + 
  168\,{  A_1}\,{{{  A_2}}^3}\,{  B_2} + 
  96\,{{{  A_2}}^3}\,{  B_1}\,{  B_2}\cr
& + 
  96\,{{{  A_2}}^2}\,{{{  B_2}}^2} + 
  780\,{  A_1}\,{{{  A_2}}^2}\,{{{  B_2}}^2} + 
  660\,{{{  A_2}}^2}\,{  B_1}\,{{{  B_2}}^2} + 
  264\,{  A_2}\,{{{  B_2}}^3} + 2376\,{  A_1}\,{  A_2}\,{{{  B_2}}^3}\cr
& + 
  3288\,{  A_2}\,{  B_1}\,{{{  B_2}}^3} + 396\,{{{  B_2}}^4} + 
  3492\,{  A_1}\,{{{  B_2}}^4} + 6576\,{  B_1}\,{{{  B_2}}^4} + 
  48\,{  A_1}\,{{{  A_2}}^3}\,{  C_1}\cr
& + 
  432\,{  A_1}\,{{{  A_2}}^2}\,{  B_2}\,{  C_1} + 
  2496\,{  A_1}\,{  A_2}\,{{{  B_2}}^2}\,{  C_1} + 
  3552\,{  A_2}\,{  B_1}\,{{{  B_2}}^2}\,{  C_1} + 
  8160\,{  A_1}\,{{{  B_2}}^3}\,{  C_1}\cr
& + 
  20544\,{  B_1}\,{{{  B_2}}^3}\,{  C_1} + 
  8496\,{  A_1}\,{{{  B_2}}^2}\,{{{  C_1}}^2} + 
  25008\,{  B_1}\,{{{  B_2}}^2}\,{{{  C_1}}^2}\, ,
&(A5)\cr
F_{A_1C}=\,&0  \, , &(A6)\cr
F_{A_2A}=\,&
{  A_2} + 4\,{  A_1}{  A_2} + 12\,{{{  A_1}}^2}{  A_2} + 
  24\,{{{  A_1}}^3}{  A_2} + 24\,{{{  A_1}}^4}{  A_2} + 
  24\,{{{  A_1}}^2}{  A_2}{  B_1} + 
  72\,{{{  A_1}}^3}{  A_2}{  B_1}\cr
& + 
  72\,{{{  A_1}}^2}\,{  A_2}\,{{{  B_1}}^2} + 
  12\,{{{  A_1}}^2}\,{  B_2} + 48\,{{{  A_1}}^3}\,{  B_2} + 
  72\,{{{  A_1}}^4}\,{  B_2} + 72\,{{{  A_1}}^2}\,{  B_1}\,{  B_2}\cr
& + 
  288\,{{{  A_1}}^3}\,{  B_1}\,{  B_2} + 
  456\,{{{  A_1}}^2}\,{{{  B_1}}^2}\,{  B_2} + 
  528\,{  A_1}\,{{{  B_1}}^3}\,{  B_2} + 528\,{{{  B_1}}^4}\,{  B_2}\cr
& + 
  24\,{{{  A_1}}^4}\,{  C_1} + 96\,{{{  A_1}}^3}\,{  B_1}\,{  C_1} + 
  528\,{{{  B_1}}^4}\,{  C_1}\, , &(A7)\cr
F_{A_2B}=\,&12\,{{{  A_2}}^3} + 48\,{  A_1}\,{{{  A_2}}^3} + 
  72\,{{{  A_1}}^2}\,{{{  A_2}}^3} + 24\,{{{  A_2}}^3}\,{  B_1} + 
  96\,{  A_1}\,{{{  A_2}}^3}\,{  B_1} + 36\,{{{  A_2}}^2}\,{  B_2}\cr
& + 
  216\,{  A_1}\,{{{  A_2}}^2}\,{  B_2} + 
  504\,{{{  A_1}}^2}\,{{{  A_2}}^2}\,{  B_2} + 
  576\,{  A_1}\,{{{  A_2}}^2}\,{  B_1}\,{  B_2} + 
  384\,{  A_1}\,{  A_2}\,{{{  B_2}}^2}\cr
& + 
  1560\,{{{  A_1}}^2}\,{  A_2}\,{{{  B_2}}^2} + 
  2640\,{  A_1}\,{  A_2}\,{  B_1}\,{{{  B_2}}^2} + 
  1584\,{  A_2}\,{{{  B_1}}^2}\,{{{  B_2}}^2} + 
  528\,{  A_1}\,{{{  B_2}}^3}\cr
& + 2376\,{{{  A_1}}^2}\,{{{  B_2}}^3} + 
  528\,{  B_1}\,{{{  B_2}}^3} + 6576\,{  A_1}\,{  B_1}\,{{{  B_2}}^3} + 
  6576\,{{{  B_1}}^2}\,{{{  B_2}}^3}\cr
& + 
  144\,{{{  A_1}}^2}\,{{{  A_2}}^2}\,{  C_1} + 
  864\,{{{  A_1}}^2}\,{  A_2}\,{  B_2}\,{  C_1}
+ 
  2496\,{{{  A_1}}^2}\,{{{  B_2}}^2}\,{  C_1}\cr
&  + 
  7104\,{  A_1}\,{  B_1}\,{{{  B_2}}^2}\,{  C_1} + 
  10272\,{{{  B_1}}^2}\,{{{  B_2}}^2}\,{  C_1}\, ,&(A8)\cr
F_{A_2C}=\,&24\,{{{  A_2}}^5} + 240\,{{{  A_2}}^4}\,{  B_2} + 
  864\,{{{  A_2}}^3}\,{{{  B_2}}^2} + 
  2496\,{{{  A_2}}^2}\,{{{  B_2}}^3} + 7632\,{  A_2}\,{{{  B_2}}^4}\cr
& + 
  10272\,{{{  B_2}}^5} + 120\,{{{  A_2}}^4}\,{  C_1} + 
  11328\,{  A_2}\,{{{  B_2}}^3}\,{  C_1} + 25008\,{{{  B_2}}^4}\,{  C_1}
\, , &(A9)\cr
F_{B_1A}=\,&2\,{{{  A_1}}^2}{{{  A_2}}^2} + 6\,{{{  A_1}}^3}{{{  A_2}}^2} + 
  12\,{{{  A_1}}^2}{{{  A_2}}^2}{  B_1} + 
  12\,{{{  A_1}}^2}{  A_2}{  B_2} + 
  48\,{{{  A_1}}^3}{  A_2}{  B_2}\cr
& + 
  152\,{{{  A_1}}^2}{  A_2}{  B_1}{  B_2} + 
  264\,{  A_1}{  A_2}{{{  B_1}}^2}{  B_2} + 
  352\,{  A_2}{{{  B_1}}^3}{  B_2} + 
  32\,{{{  A_1}}^2}{{{  B_2}}^2}\cr
& + 140\,{{{  A_1}}^3}{{{  B_2}}^2} + 
  88\,{  A_1}{  B_1}{{{  B_2}}^2} + 
  748\,{{{  A_1}}^2}{  B_1}{{{  B_2}}^2} + 
  132\,{{{  B_1}}^2}{{{  B_2}}^2} + 
  1908\,{  A_1}{{{  B_1}}^2}{{{  B_2}}^2}\cr
& + 
  2192\,{{{  B_1}}^3}{{{  B_2}}^2} + 
  16\,{{{  A_1}}^3}{  A_2}{  C_1} + 
  352\,{  A_2}{{{  B_1}}^3}{  C_1} + 
  144\,{{{  A_1}}^3}{  B_2}{  C_1}\cr
& + 
  832\,{{{  A_1}}^2}{  B_1}{  B_2}{  C_1} + 
  3552\,{  A_1}{{{  B_1}}^2}{  B_2}{  C_1} + 
  6848\,{{{  B_1}}^3}{  B_2}{  C_1}\cr
& + 
  2832\,{  A_1}{{{  B_1}}^2}{{{  C_1}}^2} + 
  8336\,{{{  B_1}}^3}{{{  C_1}}^2}\, ,&(A10)\cr
F_{B_1B}=\,&{{{  A_2}}^4} + 4\,{  A_1}\,{{{  A_2}}^4} + 
  32\,{  A_1}\,{{{  A_2}}^3}\,{  B_2} + 
  220\,{  A_1}\,{{{  A_2}}^2}\,{{{  B_2}}^2} + 
  264\,{{{  A_2}}^2}\,{  B_1}\,{{{  B_2}}^2}\cr
& + 
  88\,{  A_2}\,{{{  B_2}}^3} + 1096\,{  A_1}\,{  A_2}\,{{{  B_2}}^3} + 
  2192\,{  A_2}\,{  B_1}\,{{{  B_2}}^3} + 186\,{{{  B_2}}^4} + 
  2192\,{  A_1}\,{{{  B_2}}^4}\cr
& + 5440\,{  B_1}\,{{{  B_2}}^4} + 
  1184\,{  A_1}\,{  A_2}\,{{{  B_2}}^2}\,{  C_1} + 
  3424\,{  A_2}\,{  B_1}\,{{{  B_2}}^2}\,{  C_1}\cr
& + 
  6848\,{  A_1}\,{{{  B_2}}^3}\,{  C_1} + 
  23520\,{  B_1}\,{{{  B_2}}^3}\,{  C_1} + 
  8336\,{  A_1}\,{{{  B_2}}^2}\,{{{  C_1}}^2}\cr
& + 
  48160\,{  B_1}\,{{{  B_2}}^2}\,{{{  C_1}}^2} + 
  76800\,{  B_1}\,{  B_2}\,{{{  C_1}}^3} + 94336\,{  B_1}\,{{{  C_1}}^4}
\, , &(A11)\cr
F_{B_1C}=\,&0\, ,&(A12)\cr
F_{B_2A}=\,&{{{  A_1}}^2}\,{  A_2} + 4\,{{{  A_1}}^3}\,{  A_2} + 
  6\,{{{  A_1}}^4}\,{  A_2} + 6\,{{{  A_1}}^2}\,{  A_2}\,{  B_1}\cr
& + 
  24\,{{{  A_1}}^3}\,{  A_2}\,{  B_1} + 
  38\,{{{  A_1}}^2}\,{  A_2}\,{{{  B_1}}^2} + 
  44\,{  A_1}\,{  A_2}\,{{{  B_1}}^3} + 44\,{  A_2}\,{{{  B_1}}^4}\cr
& + 
  3\,{{{  A_1}}^2}\,{  B_2} + 14\,{{{  A_1}}^3}\,{  B_2} + 
  26\,{{{  A_1}}^4}\,{  B_2} + 32\,{{{  A_1}}^2}\,{  B_1}\,{  B_2} + 
  140\,{{{  A_1}}^3}\,{  B_1}\,{  B_2}\cr
& + 
  44\,{  A_1}\,{{{  B_1}}^2}\,{  B_2} + 
  374\,{{{  A_1}}^2}\,{{{  B_1}}^2}\,{  B_2} + 
  44\,{{{  B_1}}^3}\,{  B_2} + 636\,{  A_1}\,{{{  B_1}}^3}\,{  B_2}\cr
& + 
  548\,{{{  B_1}}^4}\,{  B_2} + 12\,{{{  A_1}}^4}\,{  C_1} + 
  72\,{{{  A_1}}^3}\,{  B_1}\,{  C_1} + 
  208\,{{{  A_1}}^2}\,{{{  B_1}}^2}\,{  C_1}\cr
& + 
  592\,{  A_1}\,{{{  B_1}}^3}\,{  C_1} + 856\,{{{  B_1}}^4}\,{  C_1}
\, , &(A13)\cr
F_{B_2B}=\,&{{{  A_2}}^3} + 6\,{  A_1}\,{{{  A_2}}^3} + 
  14\,{{{  A_1}}^2}\,{{{  A_2}}^3} + 
  16\,{  A_1}\,{{{  A_2}}^3}\,{  B_1}\cr
& + 
  32\,{  A_1}\,{{{  A_2}}^2}\,{  B_2} + 
  130\,{{{  A_1}}^2}\,{{{  A_2}}^2}\,{  B_2} + 
  220\,{  A_1}\,{{{  A_2}}^2}\,{  B_1}\,{  B_2} + 
  132\,{{{  A_2}}^2}\,{{{  B_1}}^2}\,{  B_2}\cr
& + 
  132\,{  A_1}\,{  A_2}\,{{{  B_2}}^2} + 
  594\,{{{  A_1}}^2}\,{  A_2}\,{{{  B_2}}^2} + 
  132\,{  A_2}\,{  B_1}\,{{{  B_2}}^2} + 
  1644\,{  A_1}\,{  A_2}\,{  B_1}\,{{{  B_2}}^2}\cr
& + 
  1644\,{  A_2}\,{{{  B_1}}^2}\,{{{  B_2}}^2} + 22\,{{{  B_2}}^3} + 
  264\,{  A_1}\,{{{  B_2}}^3} + 1164\,{{{  A_1}}^2}\,{{{  B_2}}^3} + 
  372\,{  B_1}\,{{{  B_2}}^3}\cr
& + 4384\,{  A_1}\,{  B_1}\,{{{  B_2}}^3} + 
  5440\,{{{  B_1}}^2}\,{{{  B_2}}^3} + 
  36\,{{{  A_1}}^2}\,{{{  A_2}}^2}\,{  C_1}\cr
& + 
  416\,{{{  A_1}}^2}\,{  A_2}\,{  B_2}\,{  C_1} + 
  1184\,{  A_1}\,{  A_2}\,{  B_1}\,{  B_2}\,{  C_1} + 
  1712\,{  A_2}\,{{{  B_1}}^2}\,{  B_2}\,{  C_1} \cr
& + 
  2040\,{{{  A_1}}^2}\,{{{  B_2}}^2}\,{  C_1} + 
  10272\,{  A_1}\,{  B_1}\,{{{  B_2}}^2}\,{  C_1} + 
  17640\,{{{  B_1}}^2}\,{{{  B_2}}^2}\,{  C_1}+ 
  1416\,{{{  A_1}}^2}\,{  B_2}\,{{{  C_1}}^2}\cr
& + 
  8336\,{  A_1}\,{  B_1}\,{  B_2}\,{{{  C_1}}^2} + 
  24080\,{{{  B_1}}^2}\,{  B_2}\,{{{  C_1}}^2} + 
  19200\,{{{  B_1}}^2}\,{{{  C_1}}^3}\, , &(A14)\cr
F_{B_2C}=\,&4\,{{{  A_2}}^5} + 36\,{{{  A_2}}^4}\,{  B_2} + 
  208\,{{{  A_2}}^3}\,{{{  B_2}}^2} + 
  1272\,{{{  A_2}}^2}\,{{{  B_2}}^3}+ 4280\,{  A_2}\,{{{  B_2}}^4}
+ 
  5880\,{{{  B_2}}^5} \cr
& + 1416\,{{{  A_2}}^2}\,{{{  B_2}}^2}\,{  C_1}  + 
  8336\,{  A_2}\,{{{  B_2}}^3}\,{  C_1} + 
  24080\,{{{  B_2}}^4}\,{  C_1}\cr
& + 57600\,{{{  B_2}}^3}\,{{{  C_1}}^2} + 
  94336\,{{{  B_2}}^2}\,{{{  C_1}}^3}
\, , &(A15)\cr
F_{C_1A}=\,&{{{  A_1}}^4}\,{  A_2} + 4\,{{{  A_1}}^3}\,{  A_2}\,{  B_1} + 
  22\,{  A_2}\,{{{  B_1}}^4} + 6\,{{{  A_1}}^4}\,{  B_2}
 + 
  36\,{{{  A_1}}^3}\,{  B_1}\,{  B_2} + 
  104\,{{{  A_1}}^2}\,{{{  B_1}}^2}\,{  B_2}\cr
& + 
  296\,{  A_1}\,{{{  B_1}}^3}\,{  B_2} + 428\,{{{  B_1}}^4}\,{  B_2}+ 
  5\,{{{  A_1}}^4}\,{  C_1} + 472\,{  A_1}\,{{{  B_1}}^3}\,{  C_1} + 
  1042\,{{{  B_1}}^4}\,{  C_1}
\, , &(A16)\cr
F_{C_1B}=\,&2\,{{{  A_1}}^2}\,{{{  A_2}}^3} + 
  18\,{{{  A_1}}^2}\,{{{  A_2}}^2}\,{  B_2} + 
  104\,{{{  A_1}}^2}\,{  A_2}\,{{{  B_2}}^2} + 
  296\,{  A_1}\,{  A_2}\,{  B_1}\,{{{  B_2}}^2}\cr
& + 
  428\,{  A_2}\,{{{  B_1}}^2}\,{{{  B_2}}^2} + 
  340\,{{{  A_1}}^2}\,{{{  B_2}}^3} + 
  1712\,{  A_1}\,{  B_1}\,{{{  B_2}}^3} + 
  2940\,{{{  B_1}}^2}\,{{{  B_2}}^3}\cr
& + 
  708\,{{{  A_1}}^2}\,{{{  B_2}}^2}\,{  C_1} + 
  4168\,{  A_1}\,{  B_1}\,{{{  B_2}}^2}\,{  C_1} + 
  12040\,{{{  B_1}}^2}\,{{{  B_2}}^2}\,{  C_1} \cr
&+ 
  28800\,{{{  B_1}}^2}\,{  B_2}\,{{{  C_1}}^2} + 
  47168\,{{{  B_1}}^2}\,{{{  C_1}}^3}
\, , &(A17)\cr
F_{C_1C}=\,&{{{  A_2}}^5} + 236\,{{{  A_2}}^2}\,{{{  B_2}}^3} + 
  1042\,{  A_2}\,{{{  B_2}}^4} + 2408\,{{{  B_2}}^5} + 
  14400\,{{{  B_2}}^4}\,{  C_1}\cr
& + 47168\,{{{  B_2}}^3}\,{{{  C_1}}^2} + 
  541568\,{{{  C_1}}^5}
\, . &(A18)\cr}
$$

\vfill
\eject

\noindent{\bf References}
\vskip 0.4cm

\item{[1]} G.~H.~Weiss and R.~J.~Rubin, Adv. Chem. Phys. {\bf 52} 363
(1983). 

\item{[2]} P.~G.~ de Gennes, Adv. Colloid Interface Sci. {\bf 27} 189
(1987). 

\item{[3]} K.~Binder and K.~Kremer, in {\it Scaling Phenomena in
Disordered Systems} ed. by R.~Pynn and A. Skjeltrop (Plenum Press, New
York, 1985).

\item{[4]} K.~ De'Bell and T.~Lookman, Rev. Mod. Phys. {\bf 65} 87 (1993).

\item{[5]} E.~Eisenriegler, K.~Kremer and K.~Binder,  J. Chem. Phys.
{\bf 77} 6296 (1982).

\item{[6]} E.~Bouchaud and J.~Vannimenus, J. Phys. (Paris) {\bf 50} 2931
(1989). 

\item{[7]} V.~Bubanja, M.~Kne\v zevi\'c and  J.~Vannimenus, 
J. Stat. Phys. {\bf 71} 1 (1993).

\item{[8]} S.~Kumar and  Y.~Sing, Phys. Rev. E {\bf 48} 734 (1993).

\item{[9]} S.~Kumar, Y.~Singh and D.~Dhar, J. Phys. A {\bf 26} 4835 (1993).
 
\item{[10]} I.~\v Zivi\'c, S.~Milo\v sevi\' c  and H.~E.~Stanley 
Phys. Rev. E {\bf 49} 636 (1994).

\item{[11]} S.~Elezovi\'c-Had\v zi\'c, M.~Kne\v zevi\'c, S.~Milo\v sevi\'c
and I.~\v Zivi\'c, (unpublished)

\item{[12]} D.~Dhar, J. Math. Phys. {\bf 18} 577 (1977); ibid {\bf 19} 5
(1978). 
 
\item{[13]} S.~Kumar, Y.~Singh and Y.~Joshi, J.Phys. A: Math. Gen.
{\bf 23} 2987 (1990).

\item{[14]} S.~Kumar and Y.~Sing, J.Phys. A: Math. Gen. {\bf 23} 5115
(1990). 

\item{[15]} Z.~Borjan, M.~Kne\v zevi\'c  and S.~Milo\v sevi\' c,
   {\it Physica A} {\bf 211} 155 (1994).

\item{[16]} K.~Binder, in: {\it Phase Transitions and Critical Phenomena},
vol 8. C.~Domb and J.~L.~Lebowitz, eds. (Academic Press, New York, 1985).

\vfill\eject

\noindent{\bf Figure captions}
\vskip 0.5cm

\noindent {\bf Figure~1.} First two stages in the iterative construction
of the truncated $6-$simplex lattice; the adsorbing surface is presented
by a $5-$simplex lattice (shadow region).

\noindent 
{\bf Figure~2.}  Schematic representation of the eight restricted
partition functions for an $r-$th stage 6-simplex
fractal used in the calculation of the crossover critical
exponent $\Phi$. Here $B_2$, for example,  represents a part of a  
SAW path that enters the $r$-th order hexagon at a vertex lying on
the adsorption wall, exits it at the bulk vertex, then
enters the hexagon again through a surface vertex, and goes away, finally,
via one of the remaining unoccupied surface vertices.

\noindent
{\bf Figure~3.} (a) The exact (full circles) and the approximate (open
circles) values of the crossover exponent $\Phi$ for the surface
interacting SAWs on the truncated $n-$simplex lattices ($n=3,4,5,6$). The
full lines represent the upper and the lower bounds
on the exponent $\Phi$ (see (2.21)). 
(b) The approximate values of $\Phi$ (open circles) against $1/n\ln n$, for 
$n=10, 20,\dots , 100$. The full lines provide the upper and the lower 
bounds on the exponent $\Phi$.

\noindent 
{\bf Figure~4.} Schematic sketch of 15 restricted
partition functions needed (in addition to those presented in Fig. 2.) for
a complete description of the generating function ${\hbox{\cal
C}}_{11}(x,T)$ in the case of a $6-$simplex fractal lattice. All here
presented end points of a SAW are anchored to the attractive wall (full
circles). 

\noindent 
{\bf Figure~5.} Diagrams representing  some additional restricted
partition functions, required  for the evaluation of
the critical exponents $\gamma_1$ for SAWs on the 6-simplex lattice.
The open circles denote the ends of a SAW path ending somewhere in the
lattice bulk. 

\noindent 
{\bf Figure~6.} The exact (full symbols) and approximate (open symbols)
values of the critical exponents $\gamma_1$, $\gamma_{11}$ and $\gamma_s$
against $1/n$ ($n=3,4,5,6$), for $(T>T_{\hbox{\rms a}})$ and
$(T=T_{\hbox{\rms a}})$, for the surface 
interacting SAWs on truncated $n-$simplex lattices.
The lines that connect the exact results serve merely as guides to the 
eye.

\noindent 
{\bf Figure~7.}  The approximate values of the exponents $\gamma_s$ (squares),
$\gamma_1$ (circles) and $\gamma_{11}$ (triangles) against $\ln\ln n/\ln n$
 ($10^2\le n\le 10^4$), for the desorbed phase (a) and at the adsorption
transition (b) for surface interactin SAWs on the $n$--simplex lattices.
The full lines represent the asymptotic behavior described by (3.19) and
(3.20).

\vfill
\eject
\bye

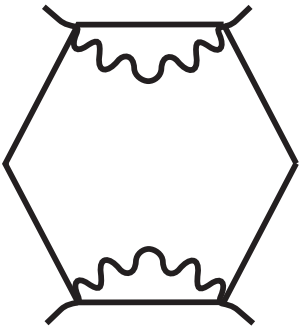
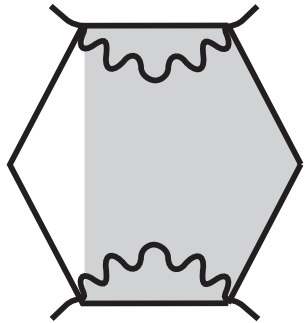
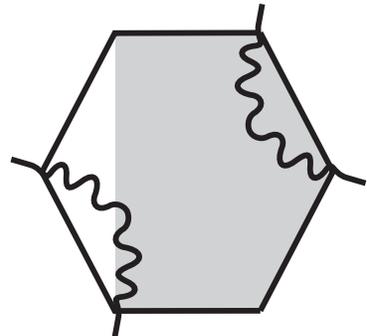
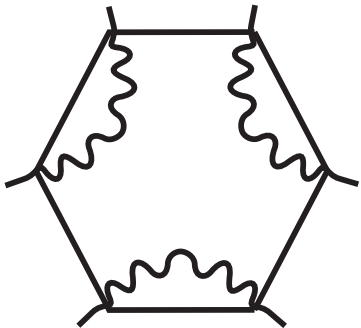
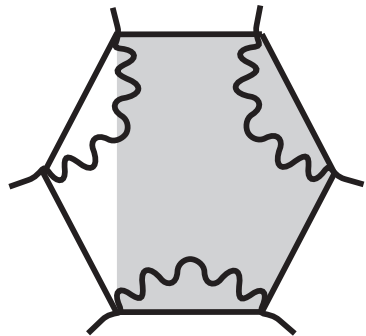
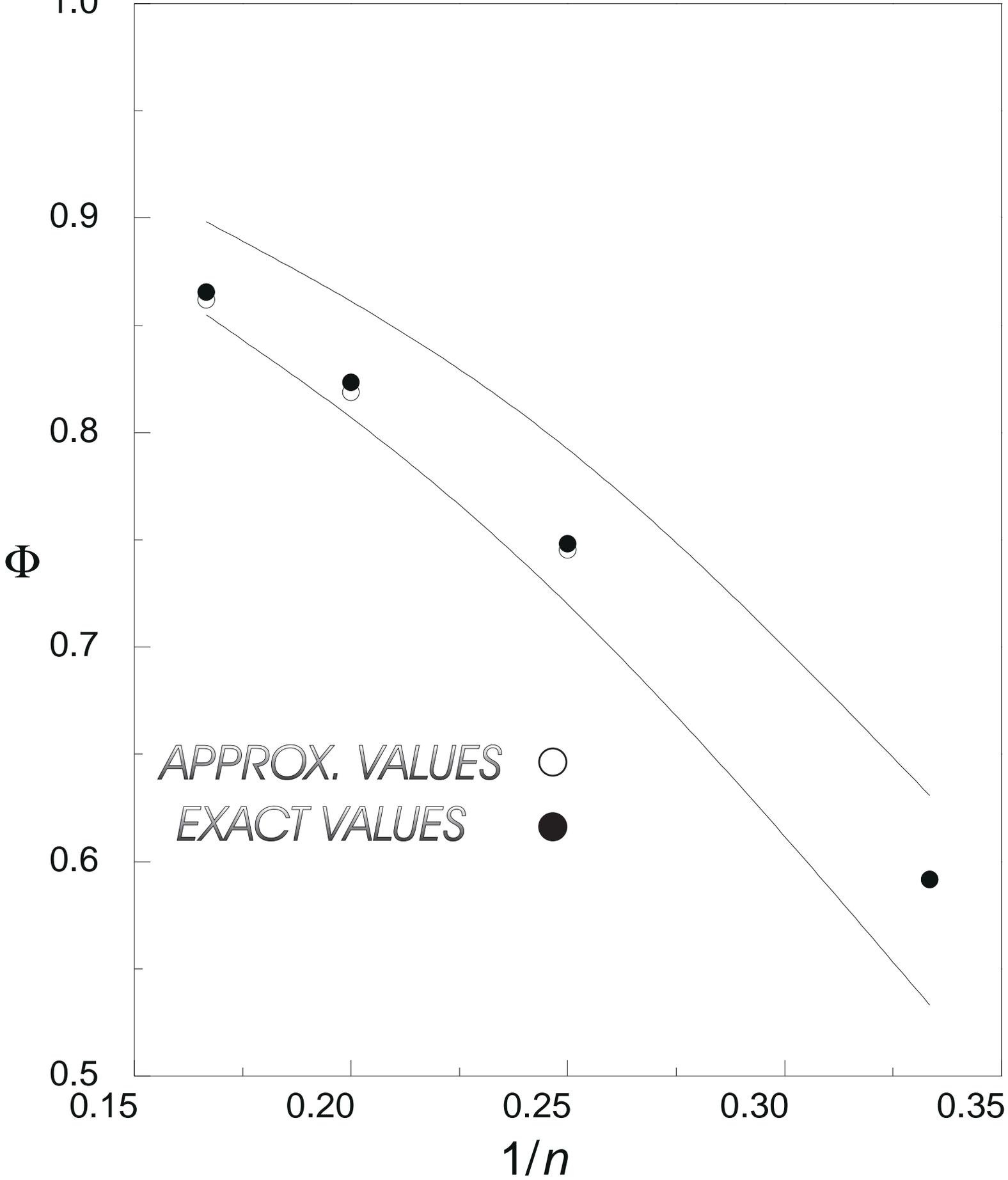
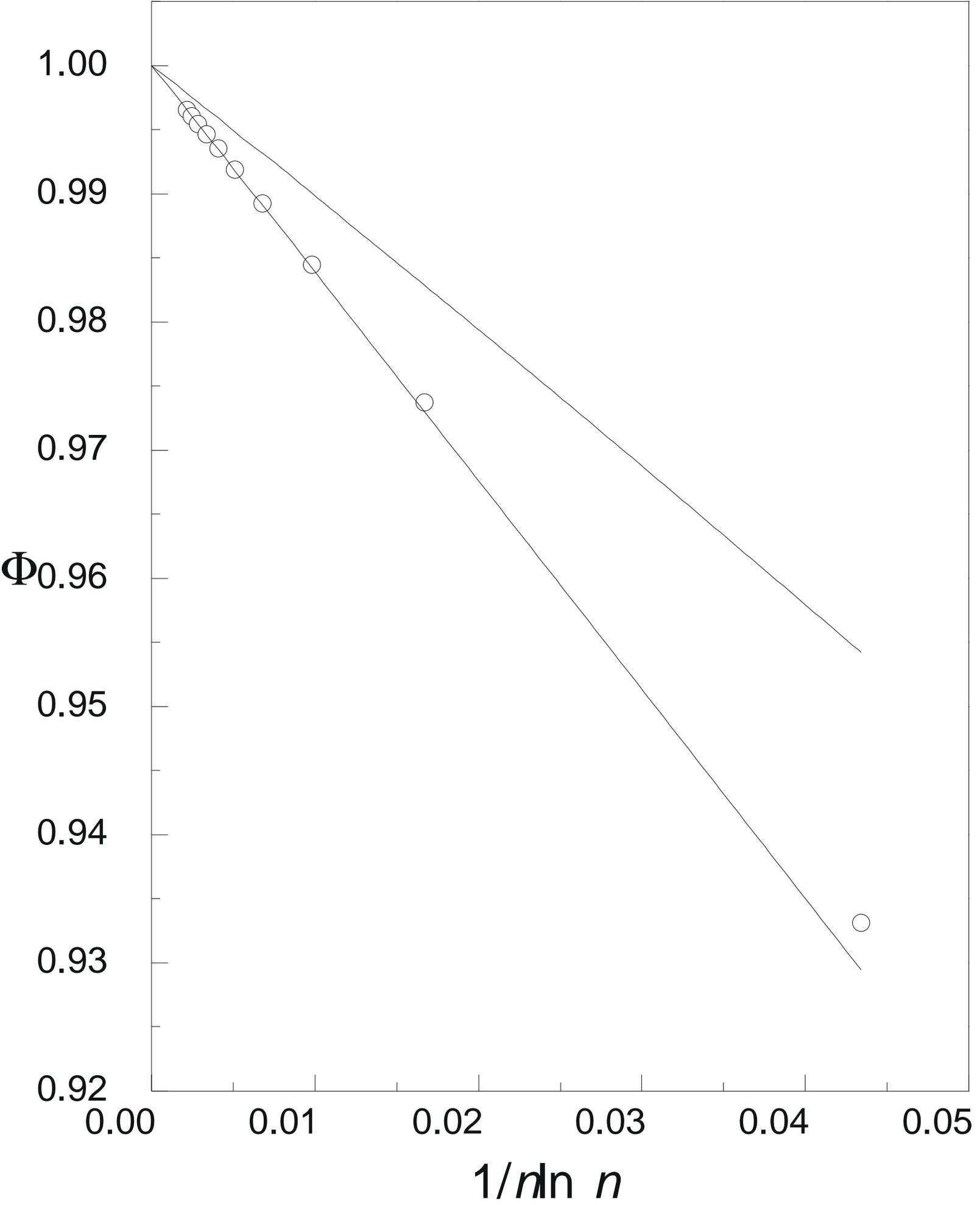
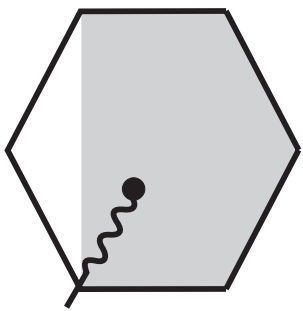
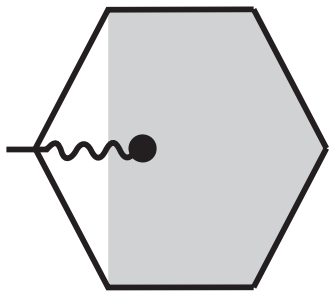
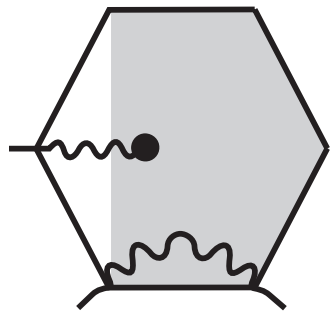
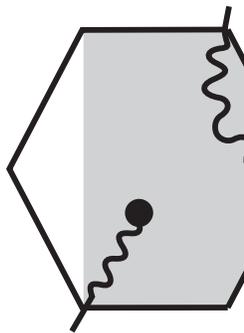
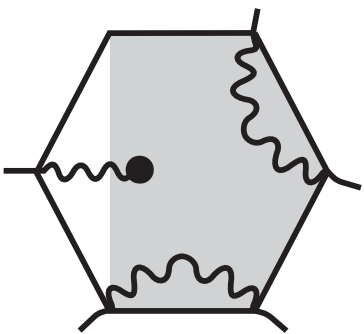
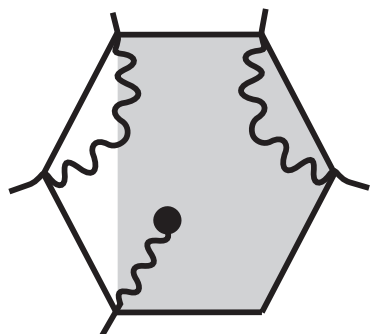
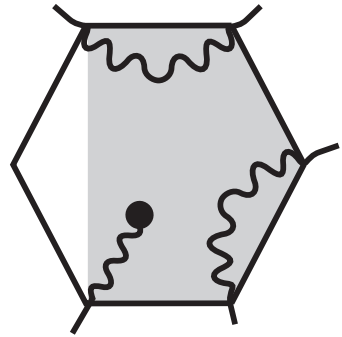
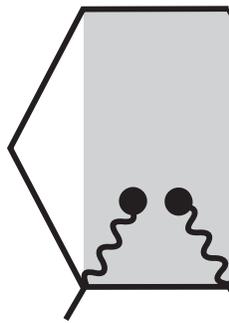
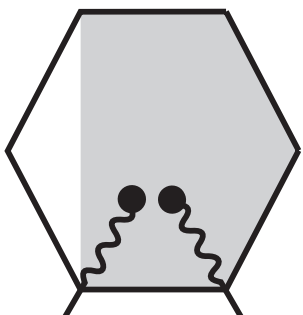
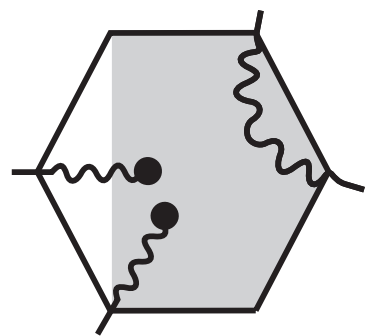
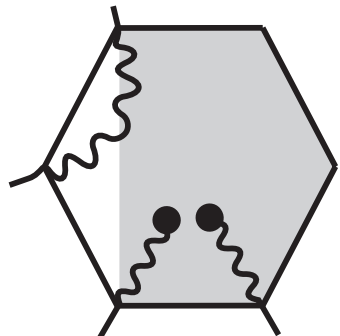
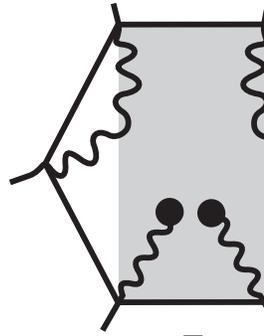
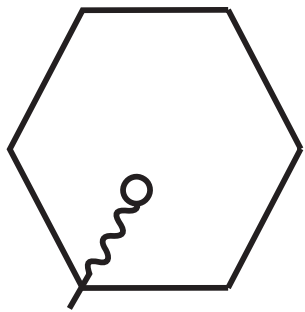
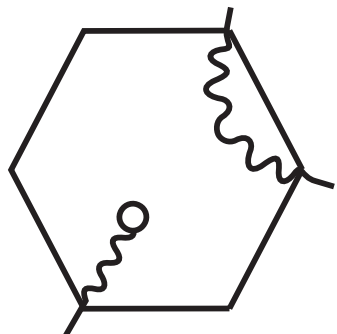
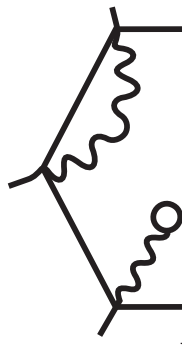
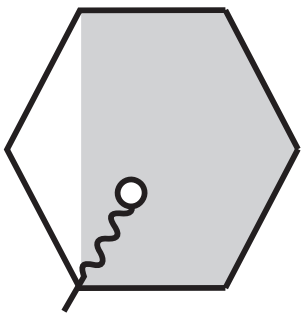
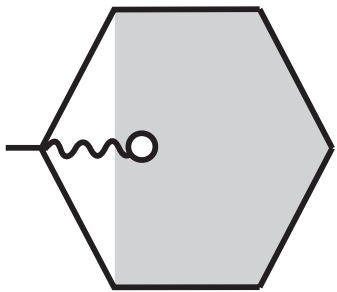
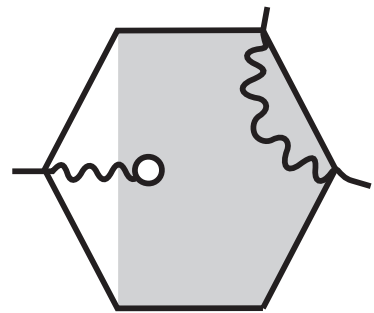
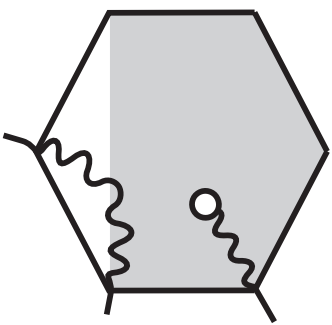
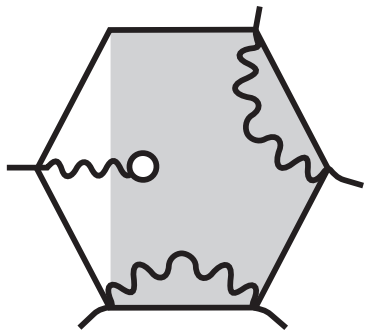
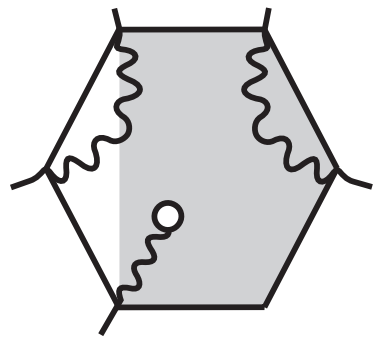
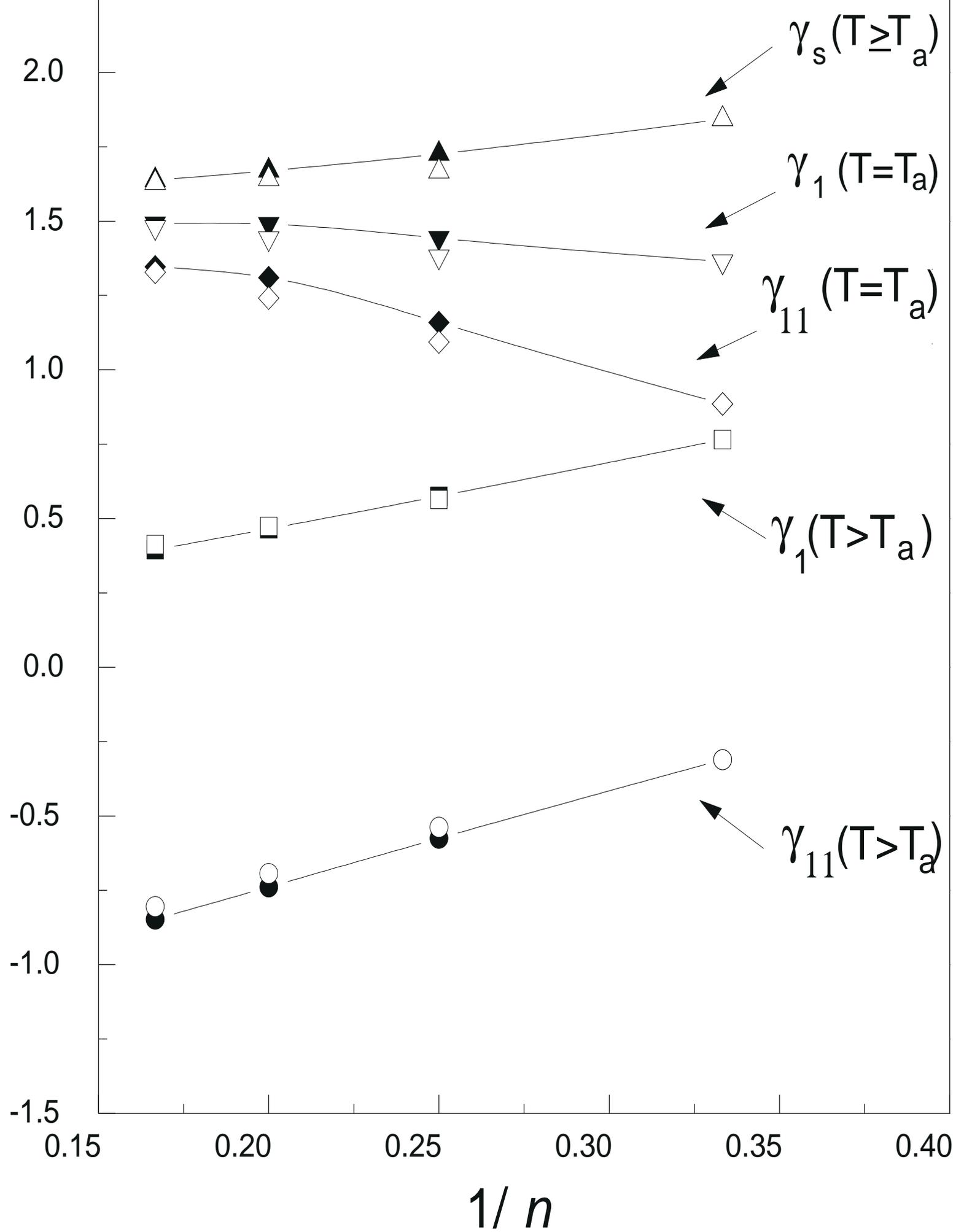
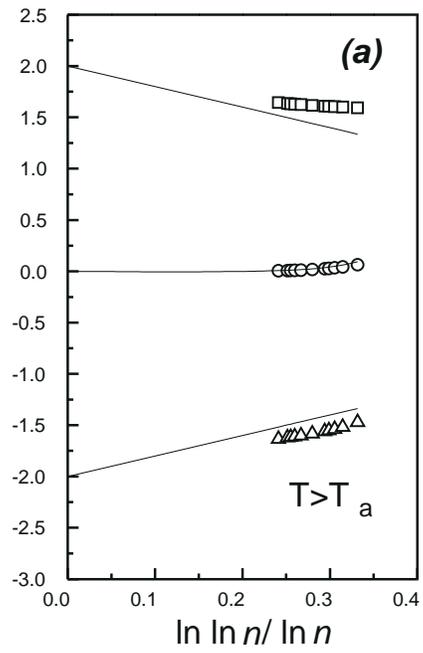
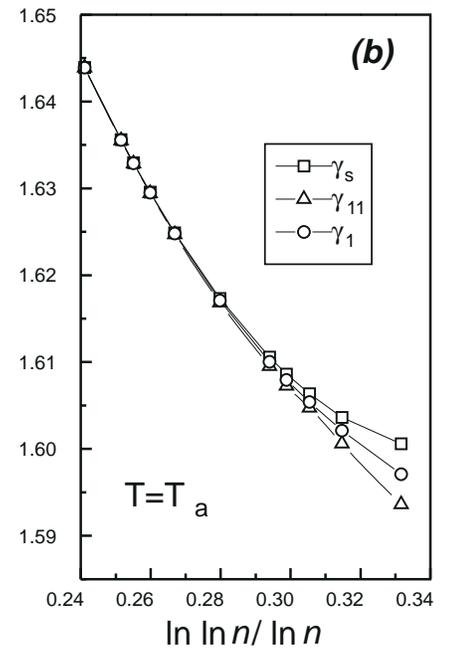
\bye